\documentclass[a4paper,11pt]{article}
\usepackage{pos}
\usepackage{amsmath}
\usepackage{braket}
\usepackage{array,multirow}
\usepackage{tikz}
\usepackage{dsfont}
\usepackage{floatrow}
\newfloatcommand{capbtabbox}{table}[][\FBwidth]

\usetikzlibrary{snakes}
\usetikzlibrary{math}
\usetikzlibrary{calc}

\newcommand{\Tab}{Table }

\tikzset{decoration={snake,amplitude=0.5mm,segment length=2mm,
		post length=0mm,pre length=0mm}}
\newcommand{\midarrow}{\tikz \draw[-stealth] (0,0) -- +(0,.1);}
\tikzset{every path/.style={line width=0.02 cm}}

\newcommand{\f}[1]{\textbf{#1}}

\newcommand{\BBdisconnected}[5] 
{
	\begin{tikzpicture}[baseline={([yshift=-.7ex]current bounding box.center)}]
	\draw[color=#4] (0,#1*1.3) -- node {\midarrow}(0,0);
	\draw[decorate, color=#5] (0,#1*1.3) to [bend left=45] (0,0);
	\draw[color=#4] (#1,1.3*#1) -- node {\midarrow} (#1,0);
	\draw[decorate, color=#5] (#1,0) to [bend left=45] (#1,#1*1.3);
	\node at (0.3*#1,1.2*#1) {#2};
	\node at (0.65*#1,1.2*#1) {#3};
	\end{tikzpicture}
}

\newcommand{\BBconnected}[5] 
{
	\begin{tikzpicture}[baseline={([yshift=-.65ex]current bounding box.center)}]
	\draw[color=#4] (0,#1*1.3) -- node {\midarrow} (0,0);
	\draw[decorate, color=#5] (0,#1*1.3) to (#1,0);
	\draw[color=#4] (#1,1.3*#1) -- node {\midarrow} (#1,0);
	\draw[decorate, color=#5] (0,0) to  (#1,#1*1.3);
	\node at (0.3*#1,1.2*#1) {#2};
	\node at (0.65*#1,1.2*#1) {#3};
	\end{tikzpicture}
}

\newcommand{\Bmeson}[5] 
{
	\begin{tikzpicture}[baseline={([yshift=-.65ex]current bounding box.center)}]
	\draw[color=#4] (0,0) -- node {\midarrow}(0,#1*1.3);
	\draw[decorate, color=#5] (0,#1*1.3) to [bend left=45] (0,0);
	\node at (0.3*#1,1.2*#1) {#2};
	\end{tikzpicture}
}

\newcommand{\ltapprox}{\raisebox{-0.5ex}{$\,\stackrel{<}{\scriptstyle\sim}\,$}}

\title{Antistatic-antistatic-light-light potentials from lattice QCD}

\author*[a]{Lasse Mueller}
\author[b]{Pedro Bicudo}
\author[c]{Marina Krstic Marinkovic}
\author[a,d]{Marc Wagner}

\affiliation[a]{Goethe-Universit\"at Frankfurt am Main, Institut f\"ur Theoretische Physik, Max-von-Laue-Stra{\ss}e 1, \\ D-60438 Frankfurt am Main, Germany}

\affiliation[b]{CeFEMA, Dep.\ F\'{\i}sica, Instituto Superior T\'ecnico, Universidade de Lisboa, Av.\ Rovisco Pais, \\ 1049-001 Lisboa, Portugal}

\affiliation[c]{Institut f\"ur Theoretische Physik, Wolfgang-Pauli-Stra{\ss}e 27, ETH Z\"urich, 8093 Z\"urich, Switzerland}

\affiliation[d]{Helmholtz Research Academy Hesse for FAIR, Campus Riedberg, Max-von-Laue-Stra{\ss}e 12, \\ D-60438 Frankfurt am Main, Germany}
\emailAdd{lmueller@itp.uni-frankfurt.de}
\emailAdd{bicudo@tecnico.ulisboa.pt}
\emailAdd{marinama@ethz.ch}
\emailAdd{mwagner@itp.uni-frankfurt.de}

\abstract{We present results for tetraquark potentials of two static anti-quarks $\bar b \bar b$ in the presence of two light quarks $u$ and/or $d$. We improve on existing results by computing the static potential also for off-axis separations, which increases the number of data points significantly. Moreover, we compute for the first time $\bar b \bar b u s$ potentials.
}

\FullConference{The 40th International Symposium on Lattice Field Theory (Lattice 2023)\\
	July 31st - August 4th, 2023\\
	Fermi National Accelerator Laboratory\\}

\begin{document}
\maketitle


\section{Introduction}

With the recent discovery of the $T_{cc}$ tetraquark at LHCb \cite{LHCb:2021auc, LHCb:2021vvq} studies of four-quark systems with two heavy anti-quarks and two light quarks have become particularly important. There have been several lattice computations in the past years studying such systems with L\"uscher's finite volume method \cite{Francis:2016hui,Francis:2018jyb,Junnarkar:2018twb,Leskovec:2019ioa,Hudspith:2020tdf,Mohanta:2020eed,Meinel:2022lzo,Padmanath:2023rdu} and the HAL-QCD method \cite{Aoki:2023nzp}. In this work we use a different approach by computing potentials between two static anti-quarks in the presence of two light quarks (for previous related work see e.g.\ Refs.\ \cite{Wagner:2010ad,Bicudo:2015kna}). Such potentials provide insights concerning the possible formation of tetraquarks and allow to study their existence and properties in the Born-Oppenheimer-approximation. For now, such potentials have only been used to study $\bar b \bar b q q$ tetraquarks \cite{Bicudo:2012qt,Brown:2012tm,Bicudo:2015vta,Bicudo:2016ooe,Bicudo:2017szl,Hoffmann:2022jdx}, but they could also be used to investigate $\bar b \bar c q q$. In the future one might even consider investigating the $T_{cc}$ tetraquark, which would, however, require additional computations of relativistic corrections (see e.g.\ Refs.\ \cite{Pineda:2000sz, Eichberg:2023trq}, where such corrections are discussed for the ordinary static potential).
Static potentials corresponding to the similar system of a static and a light quark anti-quark pair has been studied in the context of string breaking \cite{Bali:2005fu, Bulava:2019iut} and used with the Born-Oppenheimer approximation \cite{Bicudo:2019ymo, Prelovsek:2019ywc, Bicudo:2020qhp, Bicudo:2022ihz, TarrusCastella:2022rxb}. 
In this work we significantly improve on existing results \cite{Bicudo:2015kna} for the $\bar b \bar b u d$ system and we present for the first time results for the $\bar b \bar b u s$ system.


\section{Creation operators and correlation functions}
	
	We use creation operators with two anti-$b$ quarks and two light $u/d$ quarks,
	\begin{align}
		\label{EQN001} \mathcal{O}_{BB}^{I,\Gamma}(\f{r}_1,\f{r}_2) &= \left(\mathcal{C}\Gamma\right)_{AB} \left(\mathcal{C}\tilde{\Gamma}\right)_{CD}\left(\bar b^a_C(\f{r}_1) u^a_A(\f{r}_1)\; \bar b^b_D(\f{r}_2) d^b_B(\f{r}_2) \mp (u \leftrightarrow d)\right) ,
	\end{align}
	where $\mp$ corresponds to isospin $I=0$ and $I=1$, respectively, $\mathcal{C} = \gamma_0\gamma_2$ is the charge conjugation matrix and $A, B, C, D$ denote spin and $a, b$ color indices. In the static limit the spins of the $\bar b$ quarks are irrelevant and the four choices $\tilde{\Gamma} \in \{ (1 + \gamma_0) \gamma_5 , (1 + \gamma_0) \gamma_j \}$ lead to the same correlation function. $\Gamma$ couples the spin components of the light quarks and, thus, determines the quantum numbers $\Lambda_\eta^\epsilon$ as discussed in the next section.

	Temporal correlation functions of these creation operators can be expressed in terms of propagators, 
	\begin{align}
		\mathcal{C}_{BB}^{I,\Gamma}&(\f{r}_2 - \f{r}_1,t_2 - t_1) \nonumber = \langle \Omega | \mathcal{O}_{BB}^{I,\Gamma \, \dagger}(\f{r}_1,\f{r}_2;t_2) \mathcal{O}_{BB}^{I,\Gamma}(\f{r}_1,\f{r}_2;t_1) | \Omega \rangle \propto \\
		&	\propto \bigg< \left(\gamma_0\Gamma^\dagger\gamma_0\right)_{BA} \Gamma_{CD} \bigg( \nonumber \\
		& \textrm{Tr}_c \bigg[  
		U(\f{r}_1, t_2;\f{r}_1, t_1) \left(M_q^{-1}\right)_{CA}(\f{r}_1, t_1;\f{r}_1, t_2) \bigg] \times
		\textrm{Tr}_c \bigg[ U(\f{r}_2, t_2;\f{r}_2, t_1) \left(M_q^{-1}\right)_{DB}(\f{r}_2, t_1;\f{r}_2, t_2) \bigg] \nonumber\\ 
		& \pm \textrm{Tr}_c \bigg[ 
		U(\f{r}_1, t_2;\f{r}_1, t_1) \left(M_q^{-1}\right)_{CA}(\f{r}_1, t_1;\f{r}_2, t_2) \;U(\f{r}_2, t_2;\f{r}_2, t_1) \left(M_q^{-1}\right)_{DB}(\f{r}_2, t_1;\f{r}_1, t_2) \bigg] \;\bigg)\;\bigg> \equiv \nonumber\\
		&\equiv \BBdisconnected{1}{}{}{black}{black} \pm \BBconnected{1}{}{}{black}{black} , \label{eqn:corr_diagram}
	\end{align}
	where $t_2 - t_1 > 0$, $\textrm{Tr}_c$ denotes the trace in color space, $U$ is a straight path of gauge links in temporal direction (represented by a straight line in the diagram) and $M_q^{-1}$ a light quark propagator (represented by a wiggly line).

	We normalize these correlation functions by dividing by the square of the correlation function for a single $B$ meson. The large-$t$ behavior is then given by
	\begin{align}
		\label{EQN002} \frac{\mathcal{C}_{BB}^{I,\Gamma}(\f{r},t)}{(\mathcal{C}_B(t))^2} \xrightarrow[t\rightarrow \infty]{} A \exp\left(-\left(
		V^{I,\Lambda_\eta^\epsilon}_{BB}(\f{r}) - 2m_B
		\right)t\right) ,
	\end{align}
	i.e.\ the zero point of the energy corresponds to two times the $B$ meson mass.


\section{Quantum numbers of anti-static-anti-static potentials}

	In addition to isospin $I$, our $\bar b \bar b u d$ potentials can be characterized by the following quantum numbers:
	\begin{itemize}
		\item $\Lambda = \Sigma, \Pi$: Total angular momentum with respect to the separation axis of the anti-quarks.
		\item $\eta = +,- \equiv g,u$: Behavior under parity.
		\item $\epsilon = +,-$: Behavior under reflection along an axis perpendicular to the separation axis.
	\end{itemize}
	In \Tab \ref{tab:potentials} we relate all possible independent choices for $\Gamma$ with their corresponding $\Lambda_\eta^\epsilon$ quantum numbers. 
	\begin{table}
		\centering
		\begin{tabular}{c|c|c|c|c}
			& \multicolumn{2}{c}{$I=0$} & \multicolumn{2}{c}{$I=1$} \\
			$\Gamma$ & $\Lambda_\eta^\epsilon$ & shape & $\Lambda_\eta^\epsilon$ & shape\\ 
			\hline
			$\gamma_5 + \gamma_0\gamma_5$ & $\Sigma_u^+$ & A,SS & $\Sigma_g^+$ &  R,SS \\
			1 & $\Sigma_g^-$ & A,SP  & $\Sigma_u^-$ & R,SP  \\
			$\gamma_0$ & $\Sigma_u^-$ & R,SP  & $\Sigma_g^-$ & A,SP  \\
			$\gamma_5 - \gamma_0\gamma_5$ & $\Sigma_u^+$ & A,PP  & $\Sigma_g^+$ & R,PP  \\
			\hline
			$\gamma_3 + \gamma_0\gamma_3$ & $\Sigma_g^-$ & R,SS  & $\Sigma_u^-$ & A,SS  \\
			$\gamma_3\gamma_5$ & $\Sigma_g^+$ & A,SP & $\Sigma_u^+$ & R,SP  \\
			$\gamma_0\gamma_3\gamma_5$ & $\Sigma_u^+$ & R,SP & $\Sigma_g^+$ & A,SP  \\
			$\gamma_3 - \gamma_0\gamma_3$ & $\Sigma_g^-$ & R,PP & $\Sigma_u^-$ & A,PP  \\
			\hline
			$\gamma_{1/2} + \gamma_0\gamma_{1/2}$ & $\Pi_g$ & R,SS  & $\Pi_u$ & A,SS  \\
			$\gamma_{1/2}\gamma_5$ & $\Pi_g$ & A,SP & $\Pi_u$ & R,SP  \\
			$\gamma_0\gamma_{1/2}\gamma_5$ & $\Pi_u$ & R,SP & $\Pi_g$ & A,SP  \\
			$\gamma_{1/2} - \gamma_0\gamma_{1/2}$ & $\Pi_g$ & R,PP & $\Pi_u$ & A,PP  \\
		\end{tabular}
		\caption{Quantum numbers and properties of the resulting $\bar b \bar b u d$ potentials: A = attractive, R = repulsive; SS, SP ,PP = asymptotic value $2 m_B$, $m_B + m_{B_0^\ast}$, $2 m_{B_0^\ast}$.}
		\label{tab:potentials}
	\end{table}


\section{Lattice setup}

	For our computations we resort to 100 gauge link configurations from a single ensemble generated within the CLS effort \cite{Fritzsch:2012wq,Engel:2014eea} using two dynamical flavours of $O(a)$-improved Wilson-quarks and the Wilson plaquette action. The lattice size is $(L/a)^3 \times T/a = 32^3 \times 64$ with lattice spacing $a \approx 0.0755 \, \textrm{fm}$ and pion mass $m_\pi \approx 331 \, \text{MeV}$. The static action of the $\bar b$ quarks is the HYP2 static action.

	We use stochastic timeslice propagators for the light quarks with 12 stochastic sources per timeslice on 8 timeslices per gauge link configuration.

	We crudely optimize the ground state overlaps generated by our creation operators (\ref{EQN001}) by using Gaussian smearing for the quark fields with APE smeared spatial links. We performed $N_{\textrm{Gauss}} = 50$ steps of Gaussian smearing with $\kappa = 0.5$ and $N_{\textrm{APE}}=30$ steps of APE-smearing with $\alpha_{\textrm{APE}} = 30$ (With the smearing algorithms and notation consistent with \cite{Jansen:2008si}). 

	Our computations were done using the openQ*D codebase \cite{Campos:2019kgw}.


\section{Results for $\bar b \bar b u d$ potentials}
	
	\begin{figure}
		\newcommand{\scale}{0.48}
		\centering
		\includegraphics[width=\scale\textwidth]{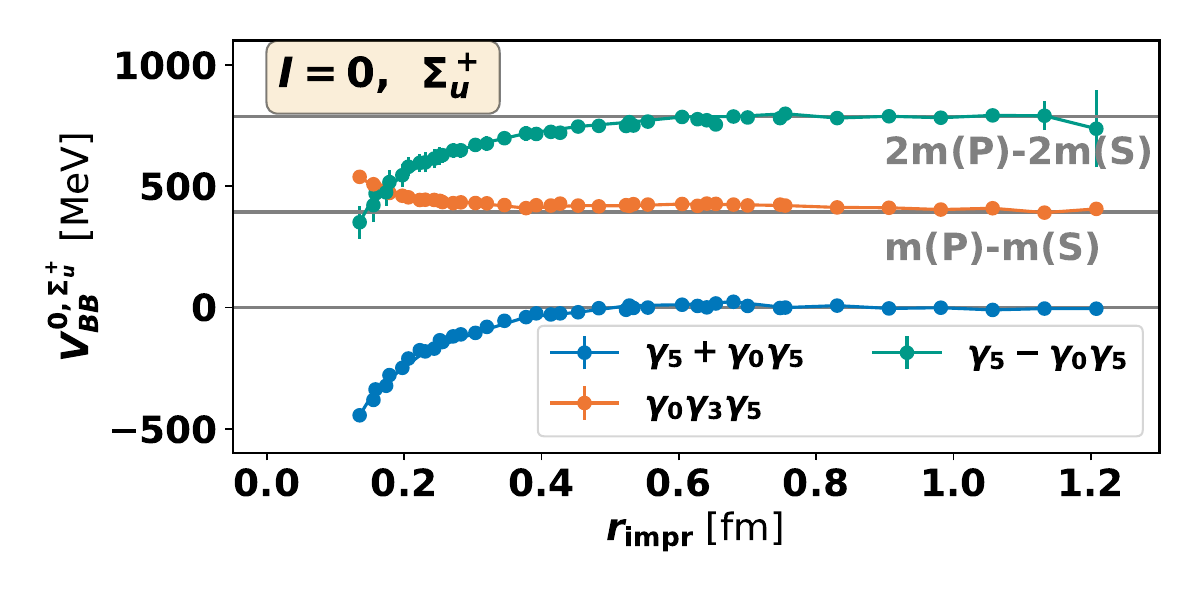}
		\includegraphics[width=\scale\textwidth]{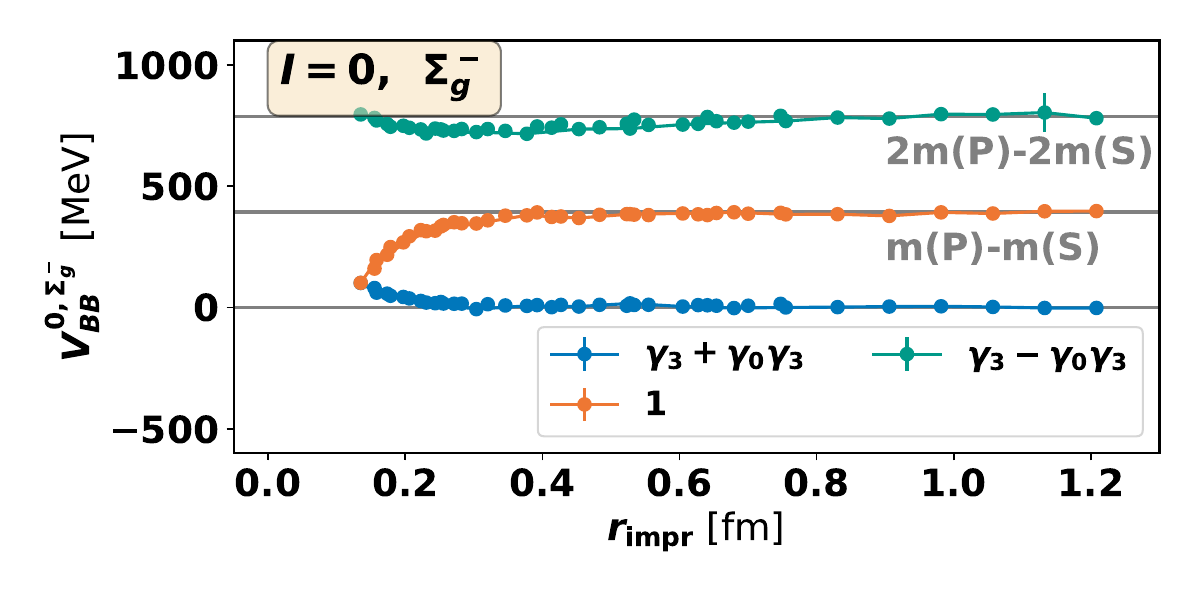}
		\includegraphics[width=\scale\textwidth]{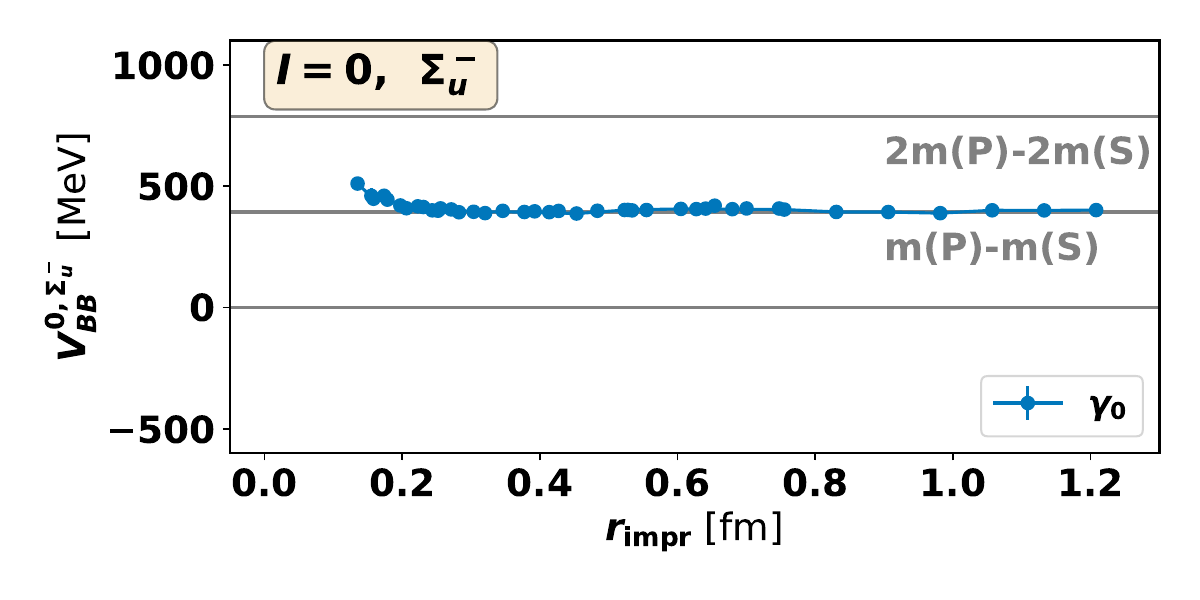}
		\includegraphics[width=\scale\textwidth]{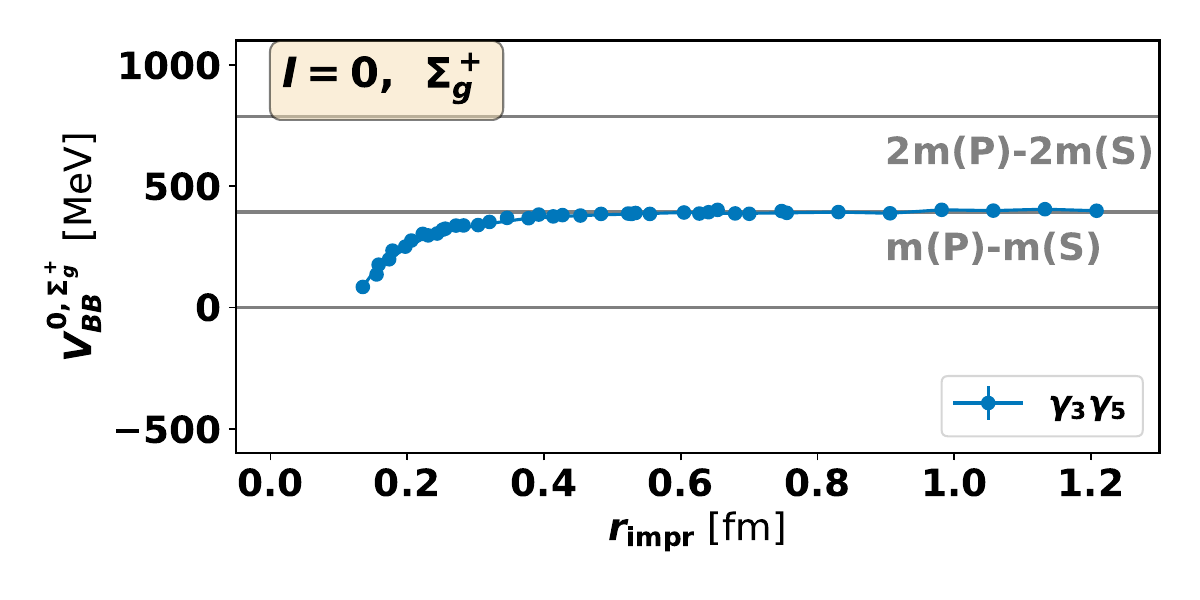}
		\includegraphics[width=\scale\textwidth]{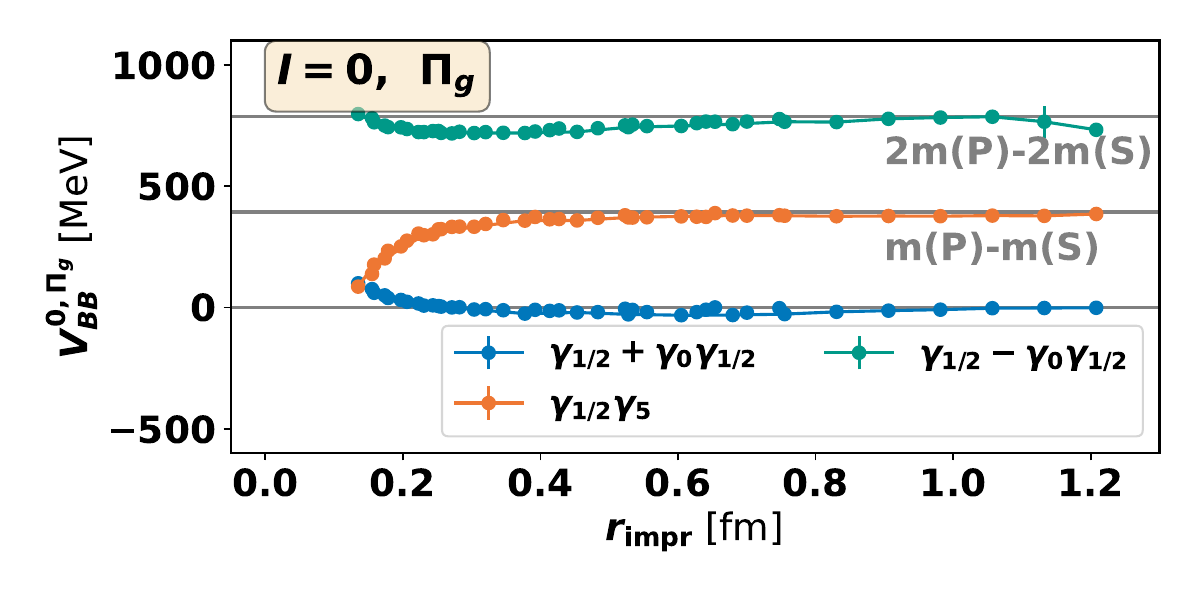}
		\includegraphics[width=\scale\textwidth]{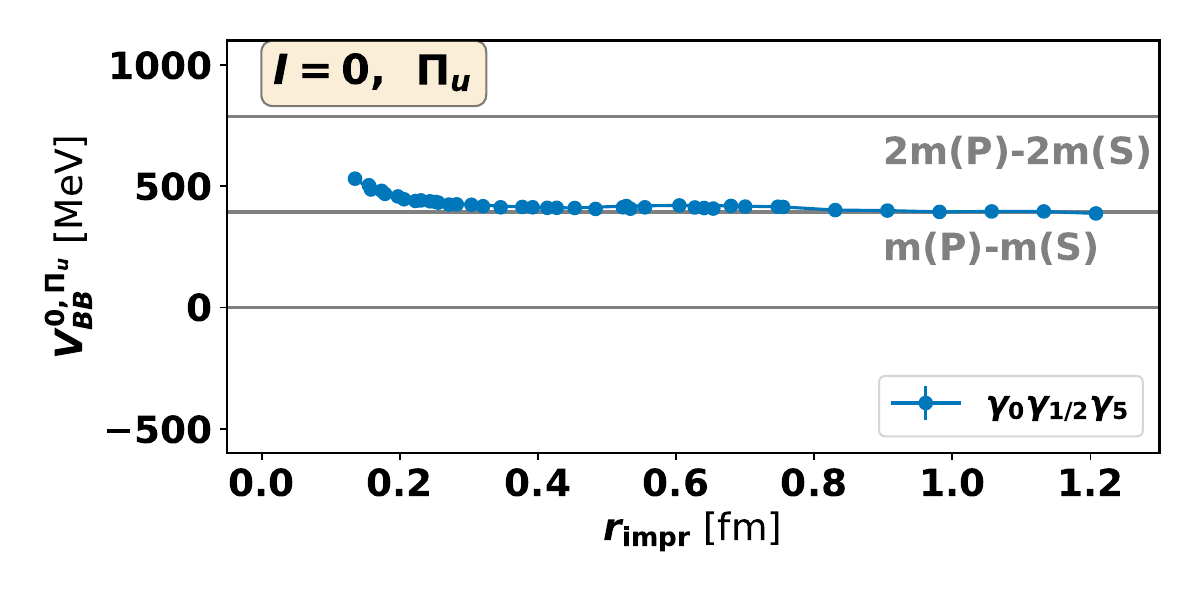}
		\includegraphics[width=\scale\textwidth]{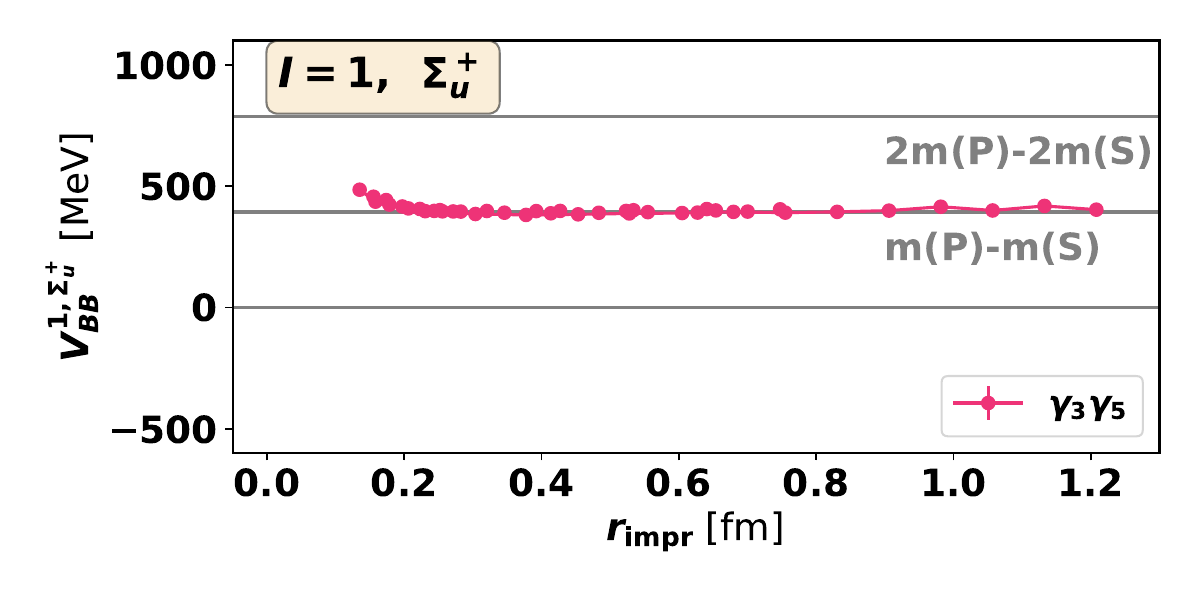}
		\includegraphics[width=\scale\textwidth]{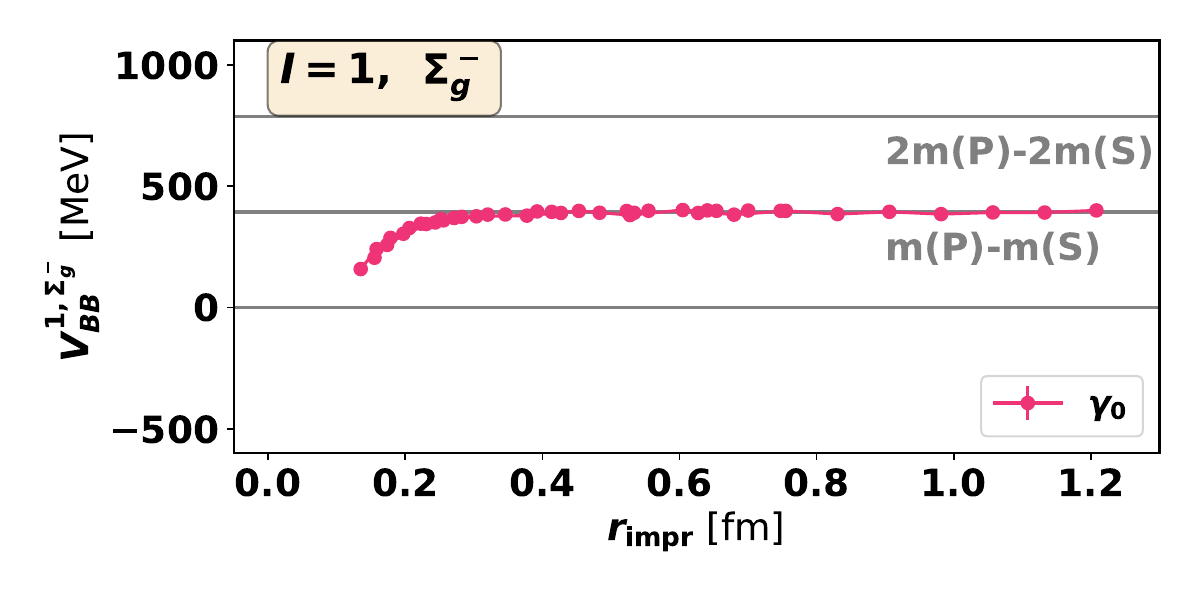}
		\includegraphics[width=\scale\textwidth]{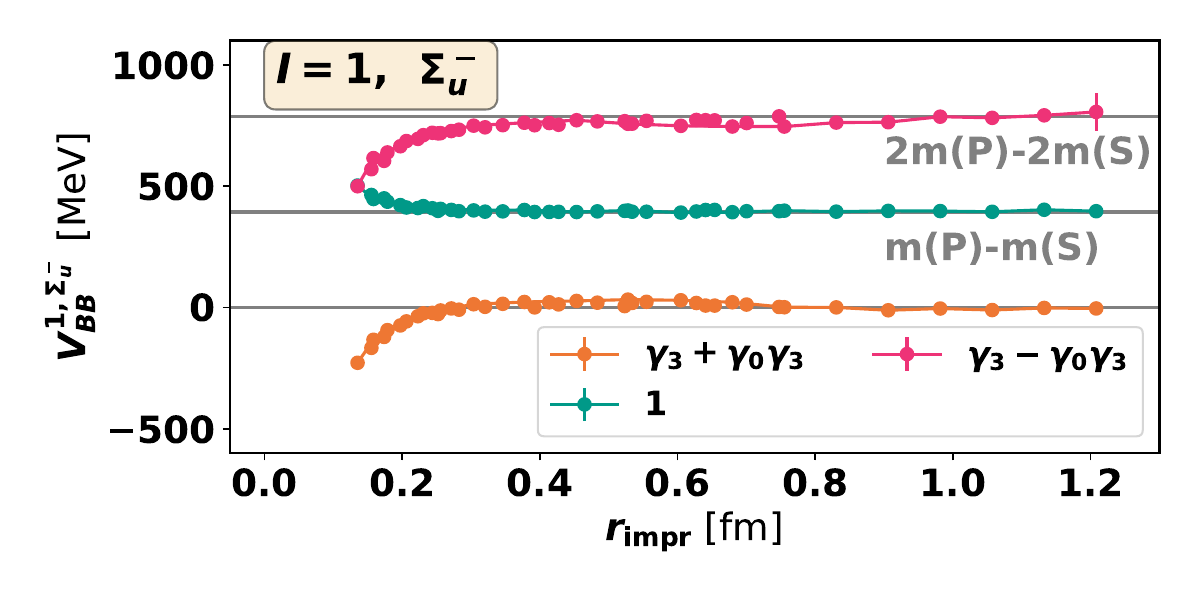}
		\includegraphics[width=\scale\textwidth]{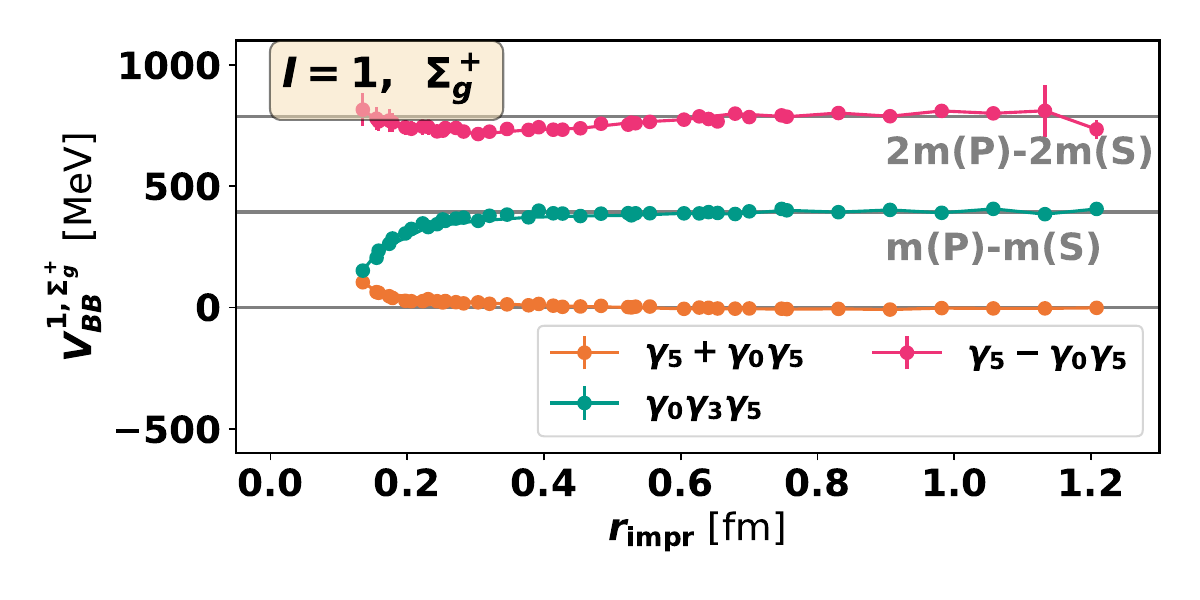}
		\includegraphics[width=\scale\textwidth]{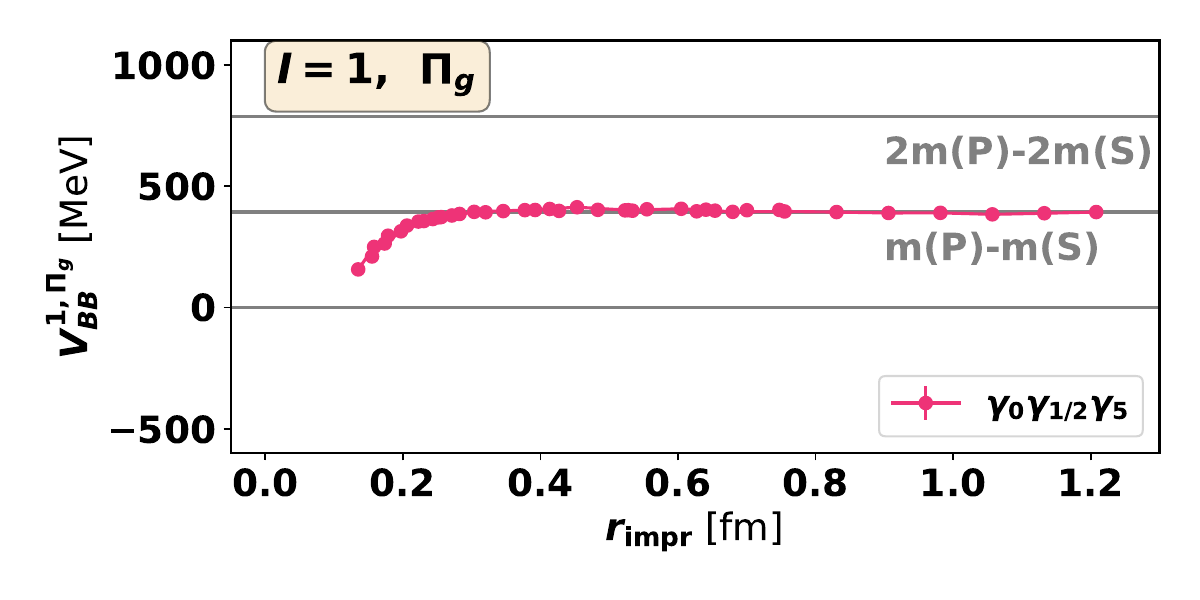}
		\includegraphics[width=\scale\textwidth]{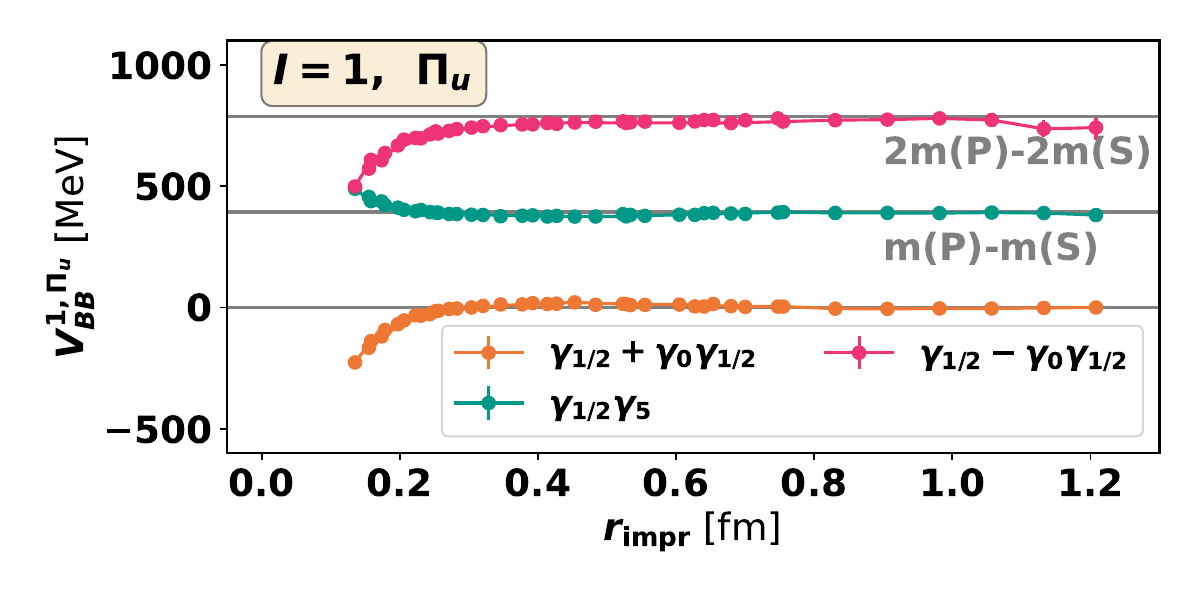}
		\caption{\label{fig:all_potentials}$\bar b \bar b u d$ potentials extracted from correlation functions corresponding to the 24 independent creation operators collected in Table~\ref{tab:potentials}. }
	\end{figure}

	We extract $\bar b \bar b u d$ potentials from the large-$t$ behavior of correlation functions (\ref{EQN002}). We consider separations $\f{r}$ along the coordinate axes up to $|\f{r}| = 16 a$. To obtain a fine spatial resolution in the interesting region of small separations, we also consider off-axis separations. We compute all possible off-axis separations for $|\f{r}| \leq 4 a$ and a subset of off-axis separations, containing the space diagonal $(x,x,x)$, the plane diagonal $(x,x,0)$ and all separations of the form $(x,1,2)$ for $|\f{r}| \leq 10 a$ (with $x\in \mathbb{Z}$ and averaging over all permutations). This yields about twenty-five additional data points. To reduce lattice artifacts, in particular at small $|\f{r}|$, we employ tree level improvement \cite{Sommer:1993ce}. In Fig.~\ref{fig:all_potentials} we show the corresponding results for the 24 independent creation operators collected in Table~\ref{tab:potentials}. Each potential is either attractive or repulsive and has one of three characteristic asymptotic values $2 m_B$, $m_B + m_{B_0^\ast}$ and $2 m_{B_0^\ast}$ ($m_B$ denotes the mass of the negative parity $B$ or $B^\ast$ meson, which are degenerate in the static limit; $m_{B_0^\ast}$ denotes the mass of the positive parity $B_0^\ast$ or $B_1^\ast$ meson, which are degenerate in the static limit and are around $400 \, \text{MeV}$ heavier than the $B$ or $B^\ast$ meson; see e.g.\ Ref.\ \cite{Jansen:2008si}). Our results are in qualitative agreement with existing results obtained around a decade ago restricted to on-axis separations \cite{Bicudo:2015kna}.

	Note that we do not consider correlation matrices, but only correlation functions with the same creation operator at both times $t_1$ and $t_2$. The consequence is that excited potentials in a given $\Lambda_\eta^\epsilon$ sector (e.g.\ the orange and green curves in the upper left plot of Fig.~\ref{fig:all_potentials}) might be contaminated by lower potentials (the blue and orange curves in the same plot). 
	
\begin{figure}
			\begin{floatrow}
				\ffigbox{%
					\includegraphics[width=0.5\textwidth]{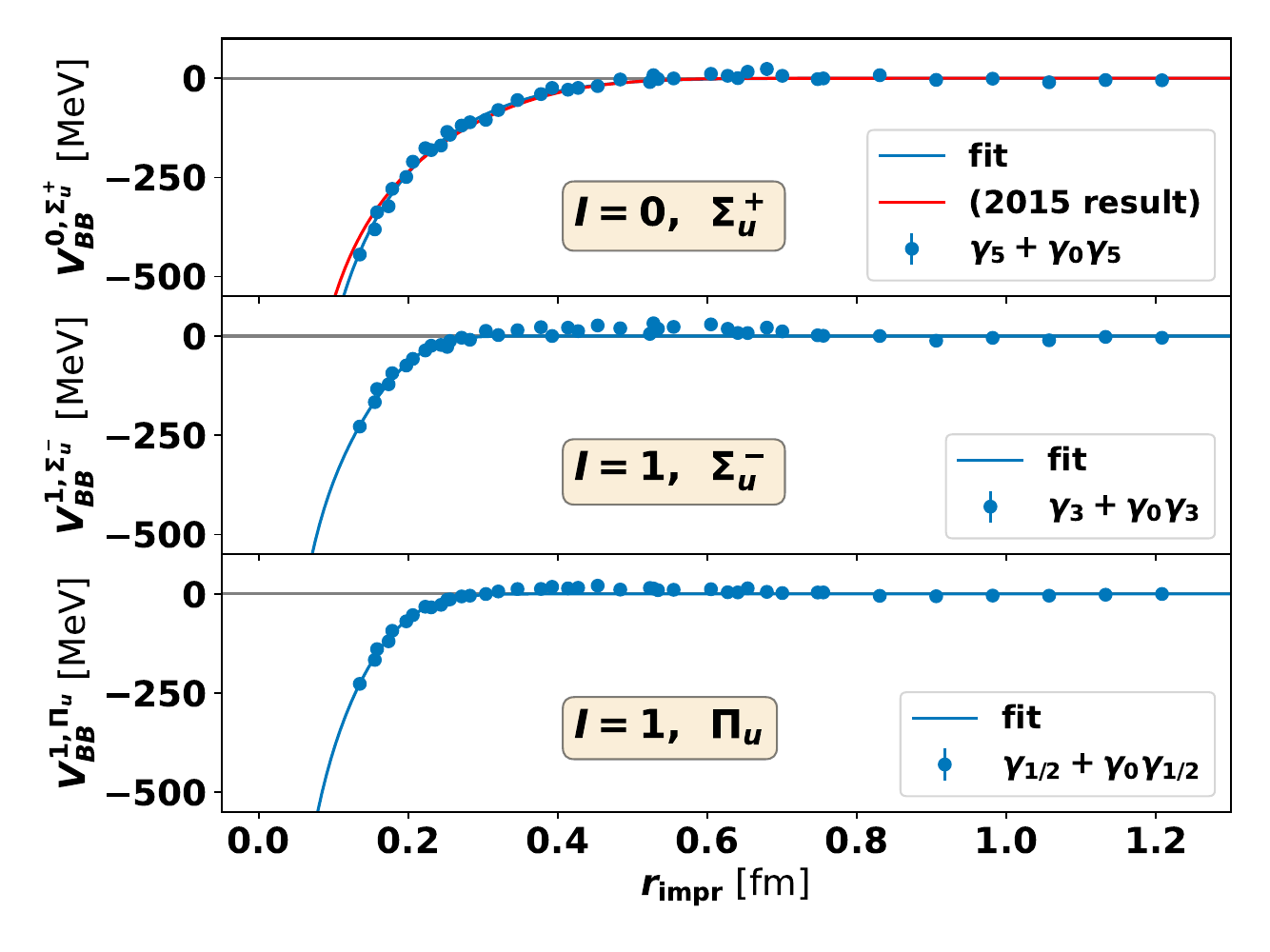}
				}{%
					\caption{Attractive $\bar b \bar b u d$ potentials with asymptotic value $2 m_B$ (blue curves represent fits with the ansatz (\ref{eqn:fit1}); the red curve is a result from Ref.\ \cite{Bicudo:2015kna}).}%
					\label{fig:groundstates_BB}
				}
				\ffigbox{%
					\includegraphics[width=0.5\textwidth]{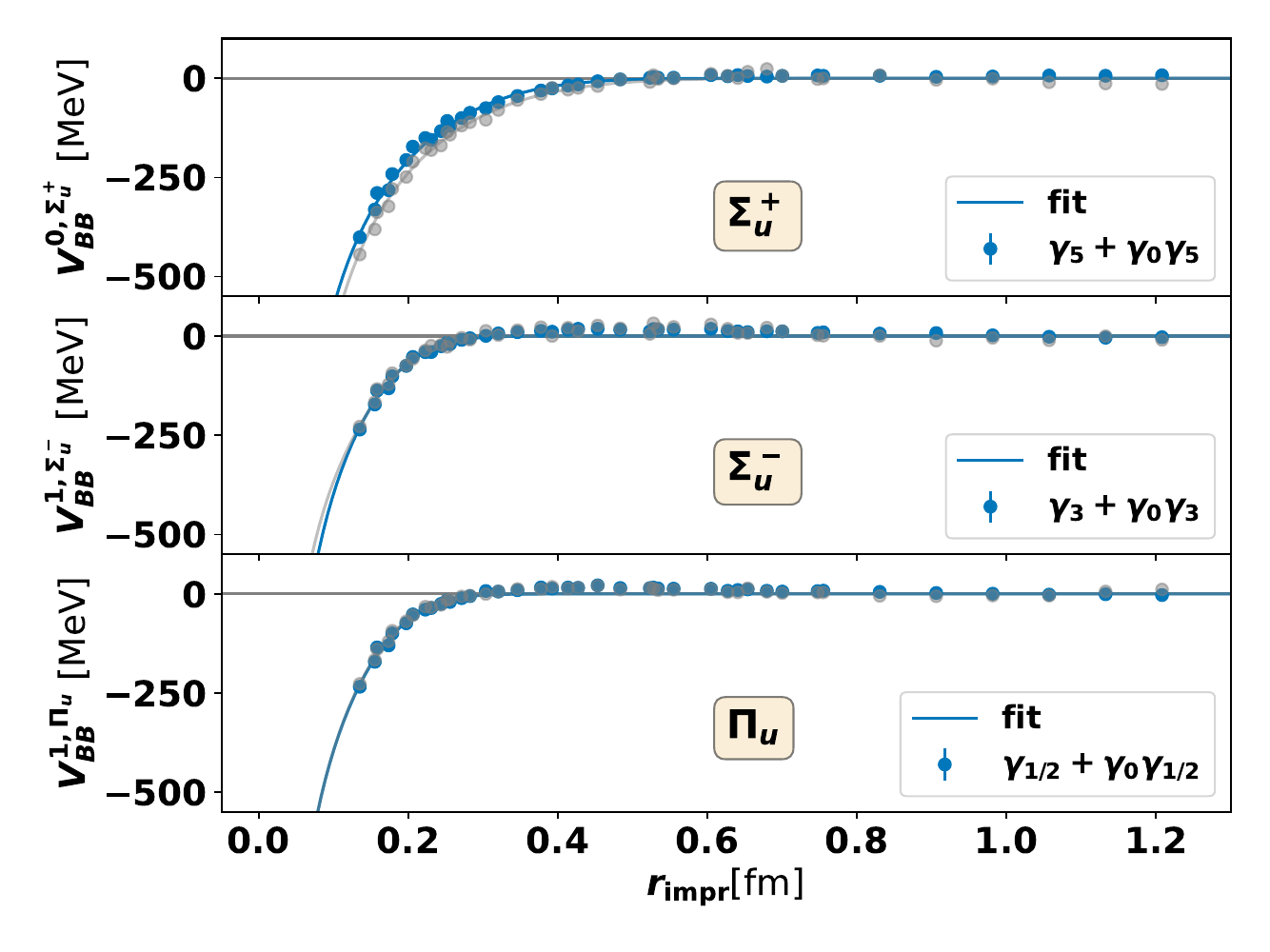}
				}{%
					\caption{Attractive $\bar b \bar b u s$ potentials with asymptotic value $m_B + m_{B_s}$ (blue curves represent fits with the ansatz (\ref{eqn:fit1})). The gray potentials represent the $\bar b \bar b u d$ results.}%
					\label{fig:groundstates_BBs}
				}
			\end{floatrow}
		\end{figure}

	From a phenomenological point of view attractive potentials with asymptotic value $2 m_B$ are of particular interest, since they are the best candidates to host bound states or resonances, which correspond to $\bar b \bar b u d$ tetraquarks. The most attractive potential has $I=0$ and $\Lambda_\eta^\epsilon = \Sigma_u^+$, while there are two related less attractive potentials with $I=1$ and $\Lambda_\eta^\epsilon = \Sigma_u^-,\Pi_u$. All three potentials can be parameterized consistently using an ansatz for a screened $1/r$ potential,
		\begin{align}
			\label{eqn:fit1} V(r) = -\frac{\alpha_1}{r}\exp \left(-\left(\frac{r}{d}\right)^p\right) .
		\end{align}
	We show these potentials with corresponding fits (blue curves) in Fig.~\ref{fig:groundstates_BB}.
	For comparison, we also show the result from Ref.\ \cite{Bicudo:2015kna} for the $I=0$ and $\Sigma_u^+$ potential obtained at similar pion mass (red curve), which is consistent with our results from this work. The $I = 1$ and $\Sigma_u^-$ and $\Pi_u$ potentials at intermediate separations $0.4 \, \text{fm} \ltapprox r \ltapprox 0.7 \, \text{fm}$ seem to be slightly above the asymptotic value $2 m_B$. Thus, there might be a weak repulsion caused by one-pion exchange. We plan to investigate this in more detail in the future by carrying out computations at smaller pion masses.

\section{Results for $\bar b \bar b u s$ potentials}

	In this work we compute for the first time also $\bar b \bar b u s$ potentials. In contrast to the $\bar b \bar b u d$ case, the two light quark propagators are different and, thus, the correlation functions contain four diagrams, 
	\begin{align}
		\mathcal{C}_{BB_s}^\Gamma(\f{r}_2-\f{r}_1,t_2-t_1) \equiv
		\BBdisconnected{1}{$u$}{$s$}{black}{black} \pm \BBconnected{1}{$u$}{$s$}{black}{black} +
		\BBdisconnected{1}{$s$}{$u$}{black}{black} \pm \BBconnected{1}{$s$}{$u$}{black}{black}.
	\end{align} 
	Due to the mass difference of the $u$ and the $s$ quark, there is only an approximate light flavor symmetry and the $I = 0$ and $I = 1$ sectors from the $\bar b \bar b u d$ case are not anymore separated for $\bar b \bar b u s$. However, the three attractive potentials with asymptotic value $m_B + m_{B_s}$ can still be extracted from their corresponding correlation functions without additional difficulties, because excited potentials with the same quantum numbers $\Lambda_\eta^\epsilon$ are around $400 \, \text{MeV}$ above. We show these $\bar b \bar b u s$ potentials in Fig.~\ref{fig:groundstates_BBs}. As expected, they are similar to their $\bar b \bar b u d$ counterparts, yet slightly less binding.
		

\section*{Acknowledgements}

	We acknowledge useful discussions with Marco Catillo, Mika Lauk, Daniel Mohler and Carolin Schlosser. 

	L.M.\ acknowledges support by a Karin and Carlo Giersch Scholarship of the Giersch foundation. 
	M.W.\ and L.M.\ acknowledge support by the Deutsche Forschungsgemeinschaft (DFG, German Research Foundation) -- project number 457742095. M.W.\ acknowledges support by the Heisenberg Programme of the Deutsche Forschungsgemeinschaft (DFG, German Research Foundation) -- project number 399217702.
	PB thanks the support of CeFEMA funded by FCT under contract UIDB/04540/2020.
	This work was performed in part at Aspen Center for Physics, which is supported by National Science Foundation grant PHY-2210452; the research was additionally supported in part by the National Science Foundation Grant No. NSF PHY-1748958.
	We acknowledge access to Piz Daint at the Swiss National Supercomputing Centre, Switzerland	under the ETHZ’s share with the project ID eth8.
	Calculations were conducted on the GOETHE-HLR and on the FUCHS-CSC high-performance computers of the Frankfurt University. We would like to thank HPC-Hessen, funded by the State Ministry of Higher Education, Research and the Arts, for programming advice.
		

\end{document}